# Impact of Oxygen Pressure on Ferroelectric Stability of La-Doped Hafnia Grown by PLD

Badari Narayana Rao,*,[†] Shintaro Yasui,[‡] Hiroko Yokota[§,∥]

[†]Center for Frontier Sciences, Chiba University, Japan.
[‡]Institute of Innovative Research, Tokyo Institute of Technology, Japan
[§]Department of Physics, Chiba University, Japan
[∥]Department of Materials Science and Engineering, School of Materials and Chemical Technology, Tokyo Institute of Technology, Japan.

**Abstract:** Thin films of $HfO_2$ doped with 4% La were fabricated on LSMO/STO (100) substrates using pulsed laser deposition. The stability of the ferroelectric orthorhombic phase in the hafnia films was investigated with respect to varying oxygen pressure during deposition. X-ray diffraction and X-ray photoelectron spectroscopy measurements were carried out to analyze the structure and composition of the films and correlated with their ferroelectric properties. Surprisingly, the ferroelectricity of the hafnia films showed a dependence on oxygen pressure during deposition of LSMO bottom electrode as well. The reason for this dependence is discussed in terms of the active role of non-lattice oxygen in the ferroelectric switching of hafnia.

## 1. Introduction

Ferroelectric materials are those that have a polar crystal structure with a spontaneous electric polarization that can be reversed by an external electric field. They are widely used in various devices, such as capacitors, pyroelectric and piezoelectric sensors and actuators, memory devices, and radio frequency and microwave devices.[1–6] While several ferroelectric materials have shown attractive properties in their bulk form, very few of them retain their ferroelectric properties in thin-film form, especially as their thickness is reduced to a few unit-cell layers.[7,8] In contrast, hafnia-based ferroelectric thin-films have recently attracted attention due to their robust ferroelectric properties even in films as thin as few unit cell layers.[9–13] Moreover, they are already being used as gate dielectrics in the semiconductor industry, and their fabrication is optimized for CMOS compatibility, making them easily integrable into existing devices.[14–17]

The ferroelectric properties of hafnia have attracted a lot of attention in recent years, and there has been a significant amount of research on these films in the past 15 years.[18–24] Several studies have investigated the effects of various factors on the fabrication of hafnia films, such as the effect of top and bottom electrode,[25–30] dopants,[19,31–34] deposition conditions,[35–39] and substrates.[37,40–43]. The most stable structure of hafnia at room temperature is the monoclinic phase, and at high temperatures, a tetragonal phase becomes more stable. Both of these phases are not ferroelectric.[44] However, it has been observed that when hafnia is cooled from high temperature, it can be stabilized in an intermediate metastable orthorhombic phase that is ferroelectric.[32,45–47] The origin of the ferroelectric structure is still under debate, and various ab-initio and experimental studies have been conducted to reveal the nature of the polar phase.[48–50] There are also reports of a ferroelectric rhombohedral phase being stabilized,[51,52] but a detailed discussion on the ferroelectric crystal structure is beyond the scope of this article. The stability of the ferroelectric phase depends on various factors, such as the substrate, dopant, electrode material, and deposition condition. In many cases, a mixed phase of orthorhombic and monoclinic or tetragonal structure is obtained.[32,36,41,53]

While most of the initial research on hafnia films was based on films deposited on silicon substrates, recently, researchers have succeeded in stabilizing the ferroelectric phase of hafnia on perovskite substrates like SrTiO$_3$, DyScO$_3$, and GdScO$_3$.[41] Estandia *et. al.* found that it was important to deposit a buffer layer of LSMO in order to stabilize the ferroelectric phase of hafnia on the perovskite substrates.[54] Nukala *et. al.* reported that LSMO acts as a source for mobile oxygen and the hafnia layer acts as oxygen vacancy acceptor or transmitter, which results in the ferroelectric switching of the film.[52] They suggested that the ferroelectric switching in these hafnia films is probably improper ferroelectricity caused due to ionic migration of oxygen between the interface.[55]

In this work, we used pulsed laser deposition (PLD) to fabricate epitaxial hafnia films on La$_{2/3}$Sr$_{1/3}$MnO$_3$ (LSMO) films, which were deposited on SrTiO$_3$ (100) (STO) single crystal substrates. While several studies have been carried out on LSMO/hafnia film heterostructures, most of the studies focused only on optimizing the growth conditions of hafnia film. We found that varying the deposition condition of LSMO also affects the ferroelectric phase stability of hafnia. In particular, we found that it was easier to obtain high-quality ferroelectric phase of hafnia on LSMO deposited at higher oxygen pressure than those deposited at lower pressures. With this consideration, we compared the structural and ferroelectric properties of 4% La-doped hafnia (4La-HFO) under different growth conditions of both hafnia as well as LSMO layers. We used La-doped HfO$_2$ for our study because researchers have shown enhanced ferroelectricity in these films and they can exhibit stable ferroelectric phase over a wide range of growth conditions.[56]

## 2. Experimental Section

LSMO films were first deposited on STO(100) single crystal substrates by PLD using the 4$^{th}$ harmonic wave of a Nd:YAG laser ($\lambda = 266$ nm) with a repetition rate of 5 Hz and laser fluence of 3 J/cm$^2$. Subsequently, 4La-HFO films were deposited on LSMO with a repetition rate of 2 Hz and laser fluence of 2 J/cm$^2$. The PLD targets were prepared by solid state synthesis, by mixing, calcining and sintering the oxides of the respective elements in appropriate ratios. For LSMO, respective oxide powders were mixed and calcined at 1200 °C for 6 hours, and pellets sintered at 1500 °C for 12 hours. For 4% La-doped HfO$_2$, respective oxide powders were mixed and calcined at 1100 °C for 6 hours, and pellets were sintered at 1400 °C for 15 hours. The STO(100) substrates were first etched in a 1:3 HNO$_3$:HCl solution and then annealed at 800 °C for two hours with a dynamic oxygen pressure of 25 Pa. LSMO was then deposited on these substrates at 775 °C and three different oxygen pressures (PO$_2$) namely 15 Pa, 20 Pa and 25 Pa. The three films are labelled as L15, L20 and L25 respectively. Similarly, for 4La-HFO deposition, the film was deposited at 700 °C and four different PO$_2$ of 1 Pa, 5 Pa, 10 Pa, and 15 Pa, and labelled as H1, H5, H10 and H15. Table 1 shows the labelling for each 4La-HFO/LSMO heterostructure deposited at different oxygen pressures. For electrical measurements, Pt-electrodes were deposited in vacuum by PLD (15 mJ, 10 Hz) on top with masks of 100 μm diameter. X-ray diffraction was carried out by using a high-resolution X-ray diffraction using a Rigaku Smartlab system using Cu-K$\alpha_1$ radiation. XPS was measured in a JEOL spectrometer (JPS-9030). Ferroelectric measurements were carried out using a Radiant ferroelectric tester (Precision LC). Out-of-plane piezoresponse force microscopy (PFM) measurements were carried out using the frequency tracking DART mode of an MFP-3D Asylum Research microscope.

| LSMO PO$_2$ (Pa) → <br> 4La-HFO (Pa) ↓ | 15 | 20 | 25 |
|---|---|---|---|
| 1 | H1L15 | H1L20 | H1L25 |
| 5 | H5L15 | H5L20 | H5L25 |
| 10 | H10L15 | H10L20 | H10L25 |
| 15 | H15L15 | H15L20 | H15L25 |

**Table 1:** Label designation for hafnia/LSMO heterostructure grown at different PO$_2$ combinations during LSMO and hafnia deposition.

## Results & Discussion

The samples were first characterized using X-ray diffraction (XRD) to obtain the structural and lattice parameter information. Figure 1 compares the out-of-plane XRD patterns of hafnia films deposited at PO$_2$ of 5 Pa, on L15, L20 and L25 LSMO films. The thickness of the films was estimated using the Laue-interference fringes observed for the Bragg peaks of the respective films.[57] The thickness of the LSMO films ranged from 45 to 60 nm [Figure S3-e], while those of 4La-HFO films ranged from 9 to 10 nm [Figure S3-d]. Peaks corresponding to hafnia appear at around 30° and 35°. While the peak at 30° corresponds to the ferroelectric orthorhombic phase, that at 35° corresponds to the non-ferroelectric monoclinic phase.[36]

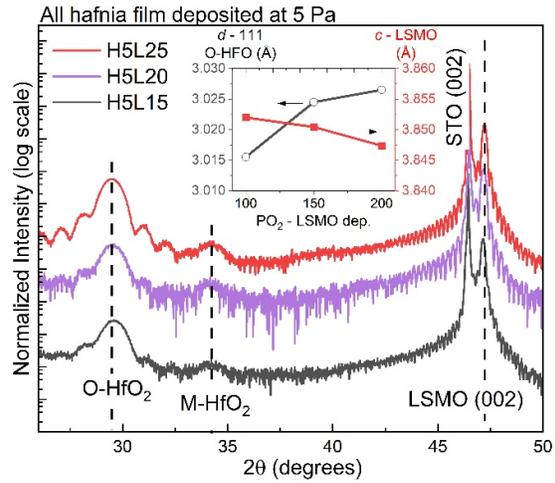

**Figure 1.** Out-of-plane XRD patterns of hafnia films, with varying PO$_2$ during LSMO deposition. Inset shows variation of interplanar $d_{(111)O-HFO}$ spacing and $c$-LSMO, as a function of LSMO deposition pressure.

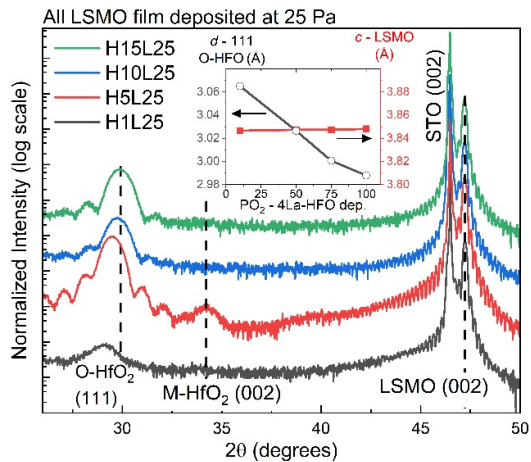

**Figure 2.** Out-of-plane XRD patterns of hafnia films, with varying PO$_2$ during 4La-HFO deposition. Inset shows variation of interplanar $d_{(111)O-HFO}$ spacing and $c$-LSMO, as a function of hafnia deposition pressure.

From figure 1, it is observed that all three patterns show a prominent orthorhombic (O-HFO) and a small monoclinic (M-HFO) peak. The H5L25 film exhibits orthorhombic peak with the highest intensity, with clearly visible Laue thickness fringes, indicating the high quality of this film. The inset in figure 1 shows that the out-of-plane $d$-spacing of LSMO decreases slightly and that of the hafnia film increases with increasing PO$_2$ during LSMO deposition. Estandia *et al.* have also shown that on a strained LSMO film, the orthorhombic phase is stabilized with decreasing out-of-plane and increasing in-plane lattice parameter of the LSMO layer.[41] However, since the decrease in the out-of-plane lattice parameter of LSMO is very small in our case (~ 0.12%), it may not be the major factor in deciding the stability of the orthorhombic phase in our films.

Next, we made a similar comparison of the XRD patterns of hafnia films deposited at different $PO_2$ on L15, L20 and L25 films. While all the films showed the presence of the orthorhombic phase of hafnia, the XRD peak intensity of the hafnia film deposited at $PO_2$ of 5 Pa (H5) was maximum for films deposited on L20 and L25, whereas the intensity monotonically increased with $PO_2$ for L15 samples [Figure S2-c]. In addition, hafnia deposited on L15 showed higher intensity monoclinic peak than other samples [Figure S2-d]. Figure 2 shows a comparison of XRD patterns of different hafnia films grown on L25, showing H5 to be the best film with high intensity O-HFO peak and clear Laue thickness fringes. In addition, we also found a monotonic decrease in the out-of-plane *d*-spacing of the orthorhombic phase with increasing $PO_2$, similar to Lyu *et. al.*[36] For hafnia samples deposited on L25, some occurrences of a small monoclinic phase peak were observed, however it was not found to have any major effect on its ferroelectric properties. In the case of H1, the orthorhombic peak is very broad and has low intensity, implying the insufficient crystallization of the orthorhombic phase.

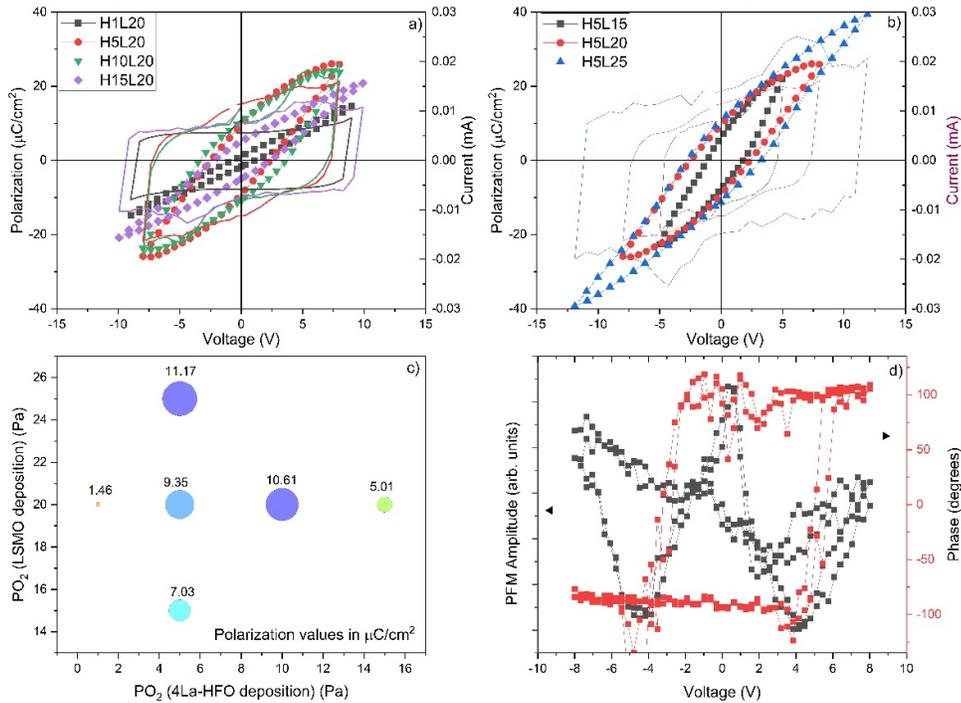

**Figure 3.** Ferroelectric properties of films grown at different $PO_2$. a) Polarization and switching current loops vs voltage with varying 4La-HFO $PO_2$. b) Polarization and switching current loops vs voltage with varying LSMO $PO_2$ c) Comparison of the remnant polarization values for all the films that showed ferroelectricity. The radius of the bubble is proportional to the magnitude of the polarization. d) Amplitude and phase loops of the H5L25 film obtained using PFM.

Figure 3 compares the ferroelectric properties of the films grown at different $PO_2$. Figure 3a compares films deposited at different $PO_2$ during hafnia depositions. We see that H1 shows paraelectric behaviour, while ferroelectricity increases for H5 and H10, and then decreases for H15. On the other hand, figure 3b compares films deposited at different $PO_2$ during LSMO deposition, and we see that hafnia deposited on L25 shows better properties than other samples.

It was also noted that the breakdown voltage of the sample improved with increasing PO$_2$ during LSMO deposition, and hence, H5L25 could sustain the highest voltage. Figure 3c shows a summary of the remnant polarization values for all the films that showed polarization switching. It should be noted that in the figure, some films have no data as they did not show any ferroelectric switching. Figure 3c clearly shows that films deposited on L20 and L25 exhibit ferroelectricity. However, the margin for change of hafnia deposition pressure is large for L20, and becomes narrow for L15 and L25 samples. On L25, although H5L25 shows the best properties, the film deposited at other pressures does not display ferroelectricity, whereas almost all the films deposited on L20 showed some ferroelectric response, even if it was smaller in magnitude. Figure 3d shows the PFM amplitude and phase measurement on H2L25, confirming ferroelectricity in the films at a local scale.

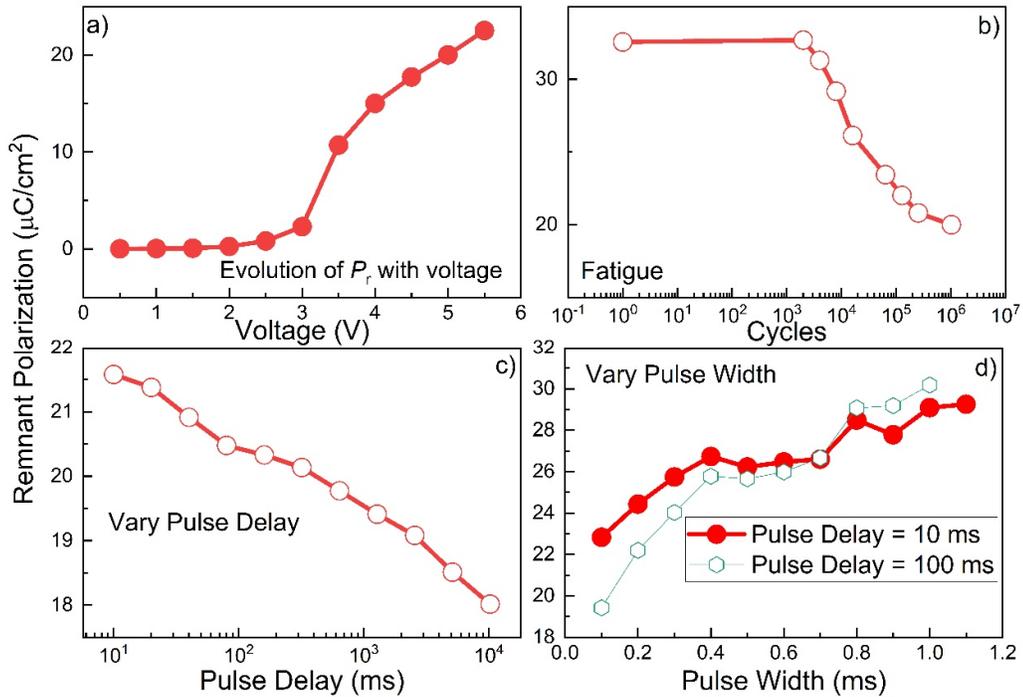

**Figure 4.** Ferroelectric properties of the optimized film H5L25. a) $P_r$ vs applied voltage, b) fatigue behaviour of $P_r$ c) $P_r$ vs pulse delay in PUND measurement, d) $P_r$ vs pulse width in PUND measurement.

Figure 4 shows a detailed characterization of the ferroelectric properties of the optimized film H5L25. Even in this optimized film, we observed a large variation in the ferroelectric properties between different Pt-pads in the film, and also there was an incessant change in the polarization values after each voltage-cycling.

Figure 5 shows the X-ray photoelectron spectra (XPS) corresponding to the oxygen 1-s shell of hafnia and LSMO layers. Typically, for hafnia, we observe two peaks arising from the oxygen. The peak at 530.2 eV corresponds to the lattice oxygen forming the Hf-O-Hf bond, and the other at 532.4 eV corresponds to the non-lattice and surface oxygen, forming sub-oxides of hafnium.[58,59] In addition, when the film is exposed to air, the moisture can form an -OH bond with the surface oxygen, which also gives rise to another peak around 531 eV.[60] In order to remove the signals from these hydroxyl bonds, the film was etched with argon inside the XPS chamber for 10 seconds. This etching was done just before measurement, so that only the peaks corresponding to hafnium oxide could be observed without any signal from hydroxyl bonds at the surface. It is important to note that the peak from non-lattice oxygen can correspond to both oxygen-deficient and oxygen-excess films. In oxygen-deficient films, the Hf-O bonds near an oxygen vacancy can change from their ideal values to compensate for the charge imbalance, thereby giving rise to the non-lattice peak in the XPS. In oxygen-excess films, the additional oxygen atoms are forced into interstitial sites, which again give rise to the non-lattice peaks.

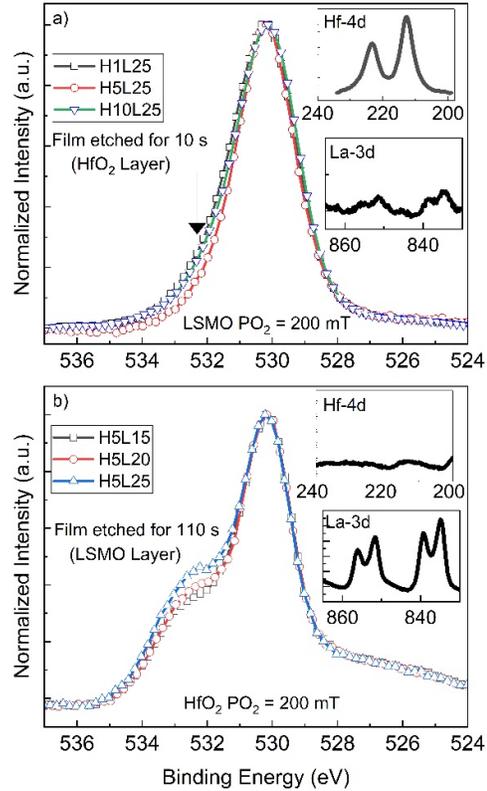

Figure 5a shows the O-1s peak from the hafnia film, and the fitting of the peaks with a Gaussian profile is given in the supporting information [Figure S4]. The calculated fraction of lattice and non-lattice oxygens is also given in the supporting information [Table S1]. For hafnia films deposited under different $PO_2$, we can see that the intensity of the secondary peak arising from non-lattice oxygen is minimum for $PO_2$ of 5 Pa (H5L25). This means that H5L25 has oxygen very close to stoichiometry, and hence, has considerably less non-lattice oxygen compared to the other films. The H1L25 film is oxygen deficient while H10L25 film is oxygen excess, both giving raise to increased intensity of the non-lattice peak in the XPS. Similar results were also observed for hafnia films deposited on L15 and L20 films (Figure S5), thereby confirming that hafnia deposited at $PO_2$ of 5 Pa gives stoichiometric film.

In figure 5b, we compare the O-1s peak on the LSMO film, after etching out the hafnia layer in the XPS chamber. The insets show the Hf-4d and La-3d XPS spectra, and the disappearance of the hafnia peak and appearance of strong La-peak confirms that all the hafnia has been etched out. Similar to the hafnia film, the O1-s spectra of LSMO film also shows a main peak corresponding to the lattice oxygen, and a secondary peak at higher binding energy corresponding to non-lattice oxygen. The O1-s spectra of the LSMO layer shows that the non-lattice peak intensity increases with increasing $PO_2$ during LSMO deposition. This indicates that LSMO has excess oxygen when deposited at higher $PO_2$ of 20Pa and 25 Pa, leading to increase in the non-lattice oxygen, which is reasonable. These weakly bonded non-lattice oxygens in LSMO may contribute to enabling the ferroelectricity in the hafnia layer.

**Figure 5.** XPS O-1s spectra for films deposited at different $PO_2$ a) 4La-HFO layer and b) LSMO layer

Nukala et. al. have carried out a comprehensive analysis using scanning transmission electron microscopy (STEM), to understand the electromigration of oxygen between the LSMO and hafnia layers during polarization switching.[52] They demonstrated that voltage dependent oxygen migration and ferroelectric switching are interrelated, and hence suggested that oxygen migration may play an important role in ferroelectric switching in hafnia. Even in our case, it appears that when an electric field is applied, some non-lattice oxygen from LSMO is injected into hafnia layer and assists in the ferroelectric switching. While a reversible migration of the non-lattice oxygen between the LSMO and hafnia films assists in the ferroelectric switching, some of these oxygen atoms can get trapped in the hafnia layer, thereby hindering the ferroelectric switching. The reason for variation in the polarization values between different voltage cycles can then be explained by considering the oxygen trapped in the hafnia layer, causing the film to show decreased polarization.

3. **Summary**

We have studied the effect of oxygen pressure on the ferroelectric stability of 4% La-doped hafnia on LSMO bottom electrode. We found that LSMO deposited at higher oxygen pressure of 20~25 Pa is more suitable to stabilize the ferroelectricity in hafnia. In addition, it was important that the oxygen is close to stoichiometry in hafnia film to obtain enhanced ferroelectric polarization. The excess oxygen in the LSMO layer is thought to reversibly migrate to the hafnia layer during application of electric field, which in turn assists in the ferroelectric switching of the hafnia film. When a part of this oxygen gets trapped in hafnia-film, it can also reduce the ferroelectric polarization.

**Author Information**

Corresponding Author

*E-mail: badari.rao@gmail.com

ORCID

Badari Narayana Rao: 0000-0003-1254-6062

Shintaro Yasui: 0000-0003-0524-9318

Hiroko Yokota: 0000-0002-2191-9257

**Acknowledgements**

Badari Narayana Rao acknowledges the financial support by the Iketani Foundation 2022 single year research grant (0341110-A), and TokyoTech collaborative research project (CRP-2022 73, CRP-2023 66). The authors gratefully acknowledge the XPS facility provided by Chiba Iodine Resource Innovation Center, Chiba University and the X-ray diffraction and AFM facility provided by Institute of Innovative Research, Tokyo Institute of Technology.

# Supporting Information

# Impact of Oxygen Pressure on Ferroelectric Stability of La-Doped Hafnia Grown by PLD

Badari Narayana Rao,*,† Shintaro Yasui,‡ Hiroko Yokota§,∥

†Center for Frontier Sciences, Chiba University, Japan.
‡Institute of Innovative Research, Tokyo Institute of Technology, Japan
§Department of Physics, Chiba University, Japan
∥Department of Materials Science and Engineering, School of Materials and Chemical Technology, Tokyo Institute of Technology, Japan.

## X-Ray Diffraction Analysis

Relative changes in orthorhombic and monoclinic peak intensity of the hafnia film are observed for different LSMO and HfO$_2$ deposition pressures. The x-ray diffraction patterns in figure S1 were taken using a simple out-of-plane diffractometer with a graphite monochromator, and hence the resolution is poor. However, the intensity variations of the orthorhombic and monoclinic phases are clearly visible.

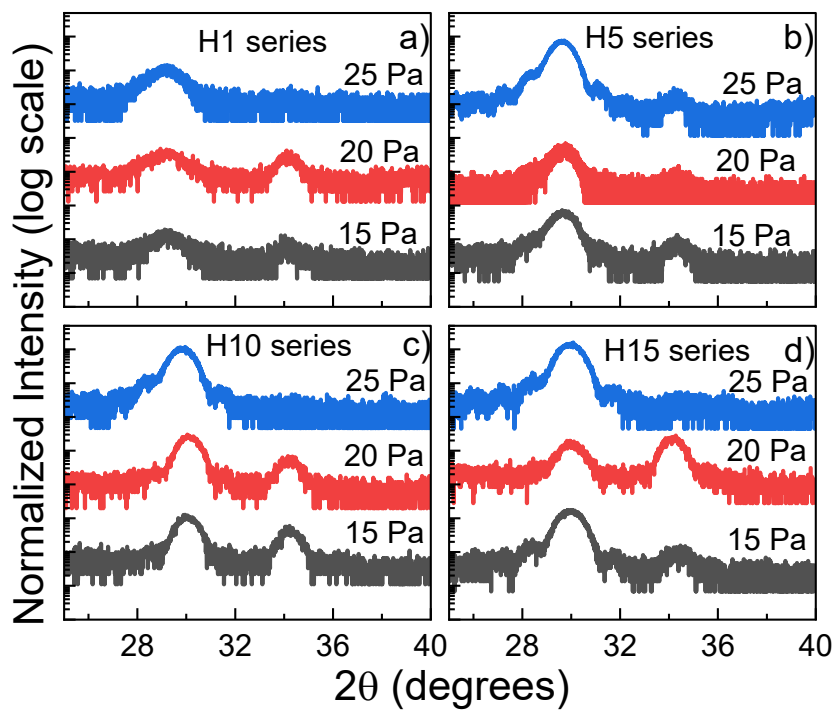

**Figure S1.** Comparison of O-HFO and M-HFO XRD peaks of films deposited under different LSMO oxygen pressures for a) H1 series, b) H5 series, c) H10 series and d) H5 series

The $d_{(111)}$ spacing of the orthorhombic phase of hafnia films decreases with increasing hafnia deposition pressure, and this trend was consistent on samples deposited on L15, L20 and L25 (Figure S2a). Similarly, the out-of-plane $c$-lattice of LSMO also decreases with increasing LSMO deposition pressure, and was observed for all H1, H5, H10 and H15 series (Figure S2b). There was a small increase in $d_{(111)}$ of O-HFO with increasing LSMO deposition pressure. In case of $c$-LSMO, it did not vary much with changing hafnia deposition pressure for L25 films, but for lower PO$_2$, c-LSMO remained minimum for H10 films, whereas it was highest for H1, and slightly higy for H5 and H15 films. The difference between *c*-LSMO was higher for L15 films than L20 films.

It can be seen that higher intensity orthorhombic peaks are obtained for higher $PO_2$ during LSMO deposition, whereas there is a majority trend for higher monoclinic phase when $PO_2$ for $HfO_2$ deposition is increased (Figure S2 c,d). However, there are exceptions to this, and it is found that when the intensity of the monoclinic peak is small, it does not have any major effect on the ferroelectric property of the film, and it only depends on the intensity of the orthorhombic peak. However, we do see reduction in ferroelectric polarization when the intensity of the monoclinic phase is large.

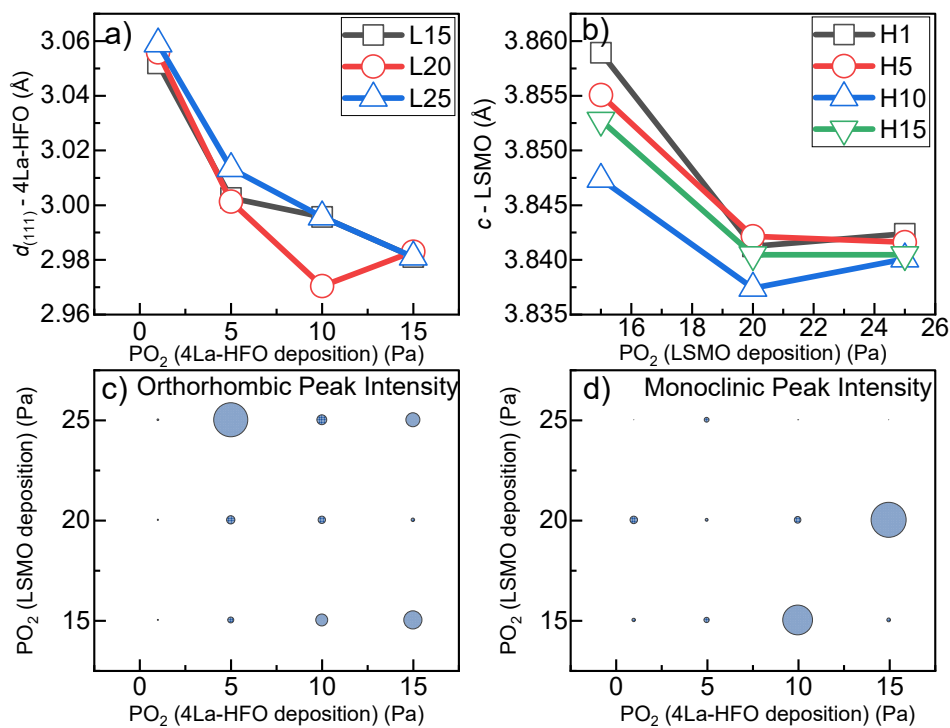

**Figure S2.** a) Comparison of $d_{(111)}$ of O-HFO with increasing HFO deposition pressure, b) comparison of *c*-LSMO with increasing LSMO deposition pressure, c) O-HFO peak intensity variation and d) M-HFO peak intensity with varying hafnia and LSMO deposition pressure. The radius of the bubbles is linearly proportional to the intensity of the corresponding XRD peaks.

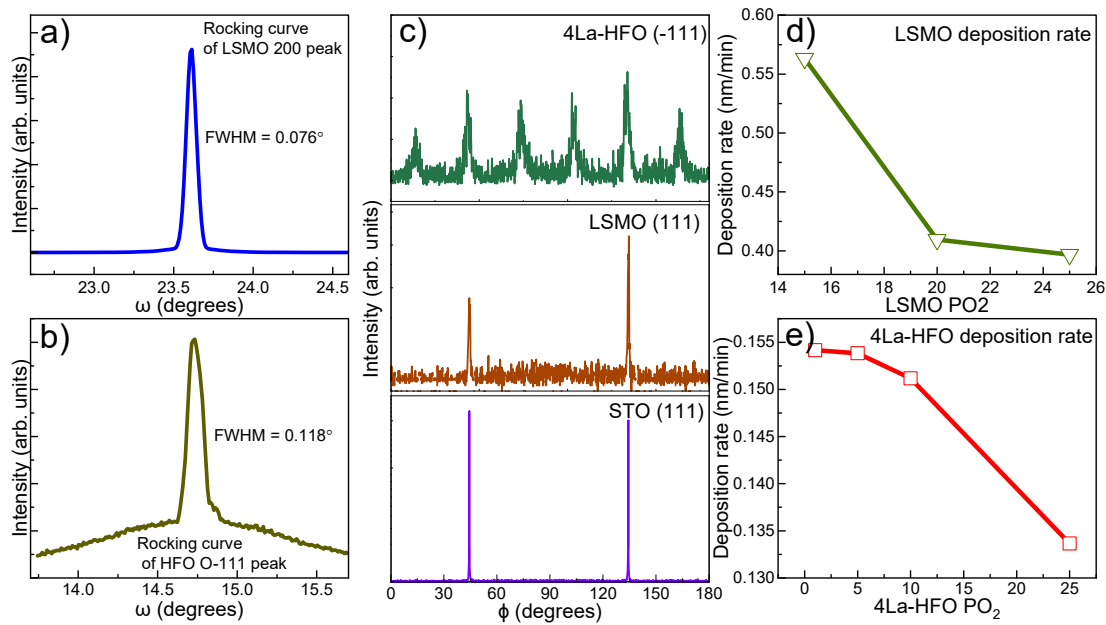

**Figure S3.** a) XRD rocking curve of LSMO 200 peak, b) XRD rocking curve of hafnia O-111 peak, c) φ-scan analysis of hafnia, LSMO and STO layers. We can confirm that LSMO has deposited epitaxially, whereas the hafnia film has six-domains in the film-plane. d) deposition rate of LSMO film. e) deposition rate of 4La-HFO film.

## X-Ray Photoelectron Spectroscopy Analysis

Comparing O-1s spectra of LSMO and hafnia layer show that for H5L25 film, hafnia has close to stoichiometric oxygen, while LSMO has large amount of non-lattice oxygen.

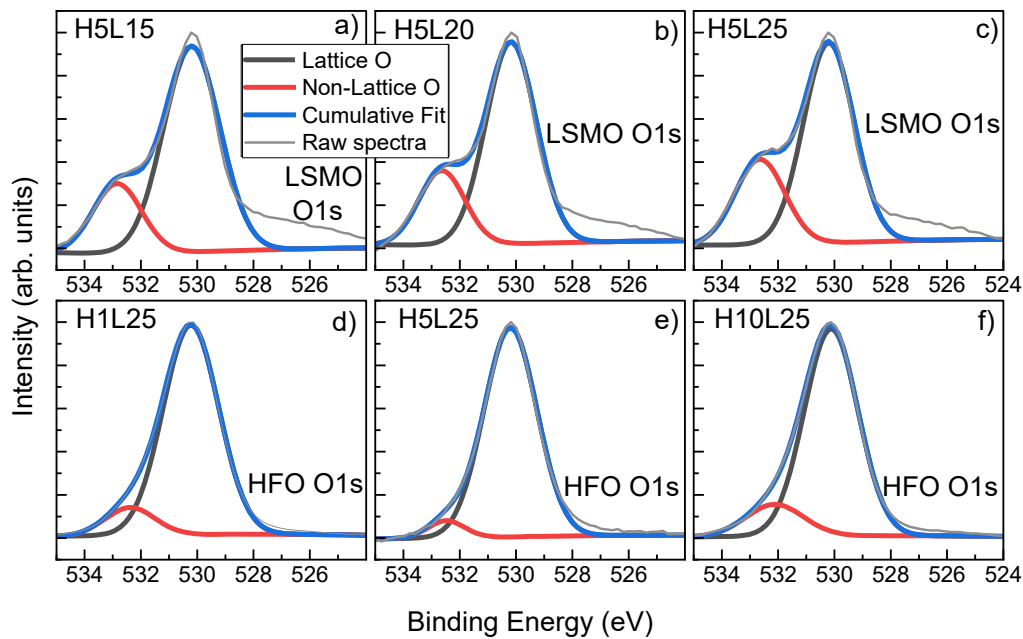

**Figure S4.** XPS O-1s spectra of (a-c) hafnia film, (d-f) LSMO film, for films deposited under different $PO_2$.

| Sample ID | Lattice O (%) | Non-Lattice O (%) |
|---|---|---|
| H1L25 | 89.4 | 10.6 |
| H5L25 | 94.6 | 5.4 |
| H10L25 | 85.9 | 14.1 |

**Table S1.** Fraction of oxygen in Lattice site and non-lattice sites for hafnia layer, as obtained from fitting in figure S4 (d-e)

| Sample ID | Lattice O (%) | Non-Lattice O (%) |
|---|---|---|
| H5L15 | 77.3 | 22.7 |
| H5L20 | 74.8 | 25.2 |
| H5L25 | 70.5 | 29.5 |

**Table S2.** Fraction of oxygen in lattice and non-lattice sites for LSMO layer, as obtained from fitting in figure S4 (a-c)

The oxygen was found to be close to stoichiometric ratio even in hafnia films deposited on L15 and L20 as seen in figure S5. This confirms that just having stoichiometric oxygen in hafnia layer is not sufficient, but we also need to have excess non-lattice oxygen in LSMO layer to observe ferroelectricity.

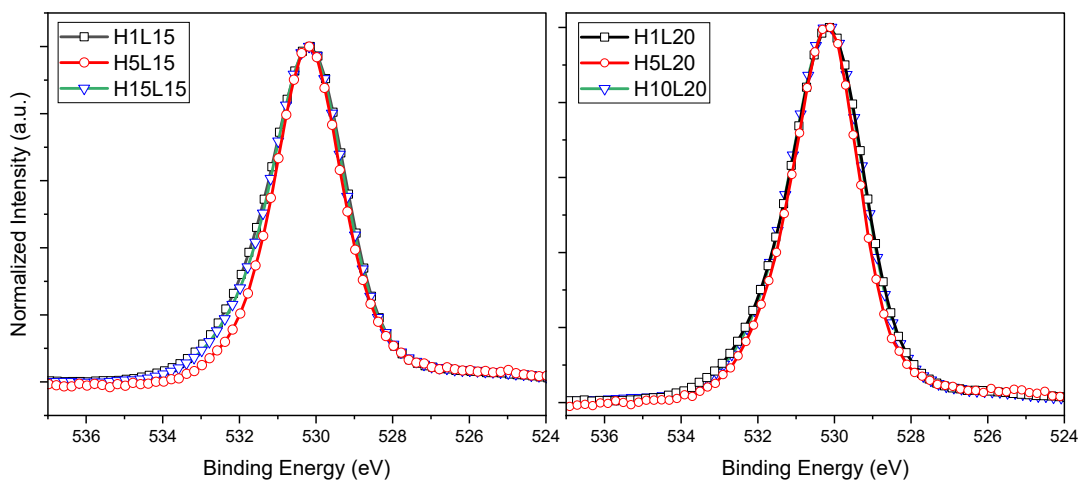

**Figure S5.** Comparison of O-1s peaks of the hafnia films deposited on L15 and L20 LSMO films. In both cases, the non-lattice peak is minimum for H5 films, thereby confirming that deposition of hafnia at 5 Pa oxygen pressure results in close to ideal oxygen stoichiometry.